%% file: main.tex
\documentclass[sigconf]{acmart}

\AtBeginDocument{%
  \providecommand\BibTeX{{%
    \normalfont B\kern-0.5em{\scshape i\kern-0.25em b}\kern-0.8em\TeX}}}

\setcopyright{acmlicensed}
\copyrightyear{2019} 
\acmYear{2019} 
\acmConference[GoodTechs '19]{EAI International Conference on Smart Objects and Technologies for Social Good}{September 25--27, 2019}{Valencia, Spain}
\acmBooktitle{EAI International Conference on Smart Objects and Technologies for Social Good (GoodTechs '19), September 25--27, 2019, Valencia, Spain}
\acmPrice{15.00}
\acmDOI{10.1145/3342428.3342678}
\acmISBN{978-1-4503-6261-0/19/09}

\usepackage{booktabs}
\usepackage{amsmath}
\usepackage[]{algorithm2e}

\begin{document}

%
\title{Open Challenges in Musical Metacreation}

\newcommand{\myuni}{University of Padua}
\newcommand{\mydept}{Department of Information Engineering}
\newcommand{\mycity}{Padua}
\newcommand{\mycountry}{Italy}
%
\author{Filippo Carnovalini}
\email{filippo.carnovalini@dei.unipd.it}
\orcid{0000-0002-2996-9486}
\affiliation{%
  \institution{\myuni}
  \institution{\mydept}
  \city{\mycity}
  \state{\mycountry}
}

%

%
\begin{abstract}
Musical Metacreation tries to obtain creative behaviors from computers algorithms composing music. In this paper I briefly analyze how this field evolved from algorithmic composition to be focused on the search for creativity, and I point out some issues in pursuing this goal. Finally, I argue that hybridization of algorithms can be a useful direction for research.
\end{abstract}

%
%

\ccsdesc[500]{Applied computing~Sound and music computing}
%
\keywords{musical metacreation, algorithmic composition, music}

%

%
\maketitle

\input{content.tex}

%
\begin{acks}
This work is funded by a doctoral grant by \myuni.
\end{acks}

%
\bibliographystyle{ACM-Reference-Format}
\bibliography{biblio,gtPHD19}

\end{document}

%% file: content.tex
\section{Introduction}
\label{sec:introduction}
Computer scientists and engineers, alongside with musicians, have used computers to make music since the earliest days of computing. This means of course the creation of digital synthesizers, but also algorithms composing pieces to be played with traditional instruments. These endeavors had many different objectives: some artists were just trying to write new music, while researchers had the aim of finding new applications for artificial intelligence. Turing himself imagined that intelligent computers could be used for the creation of musical pieces~\cite{turing_computing_2009}. 

The scientific field that is now called ``Musical Metacreation'' comprises the developments in writing computer systems capable of composing music, but the goal is actually deeper than the one of simply generating new compositions. Musical Metacreation is a subfield of ``Computational Creativity'', the discipline uniting scientist, engineers, artists, philosophers and psychologists that are trying to obtain \textit{creative} behaviors from computers~\cite{bodily_musical_2018}. Having this goal in mind many new questions arise, that can all be summed in the following: ``How can one formalize creativity?''. Such a question might seem impossible to answer, and yet many researchers have made great progress in giving scientific insights that help define creativity. Most notably, Margaret Boden with her influential book ``The Creative Mind'' described different levels of creativity by analyzing a variety of creative deeds throughout history, both coming from humans and from computers~\cite{boden_creative_2004}. From her analysis, it seems that computer systems are yet to achieve non-trivial levels of creativity.  

Despite the many advancements, it is hard to pinpoint where we are at in this research field, due to the complexity and the variety of people and systems involved. In this paper, I will try to describe some of the current issues and challenges of the field. In the conclusions, I will describe some of the directions that the field could take, and what contributions I hope to give with my research. 

\section{Creativity versus Generation}
\label{sec:creativity}

The first computer algorithms that have the aim of generating music date back to the 60s, although there are examples of algorithmic music that even predate computers. Since those years, a plethora of software for the generation of music was designed, using a variety of methods~\cite{fernandez_ai_2013}. Two main categories were distinguished: one uses a corpus of human composed music to learn a specific style, and then tries to replicate that style when composing new music. Cope's ``Experiments on Musical Intelligence'' (EMI) are a well-known example of this category~\cite{cope_computer_1992}. The other tries to generate completely new music, sometimes creating something that would not even be possible to perform for a human. 

At first, the researchers were simply trying to create some music of interest, possibly intrigued by the novelty of these applications. There is nothing wrong with that, but the most interesting questions about Computational Creativity remained unanswered. 
One might argue that Cope's algorithms were good enough in generating credible music, but the methods he used were hardly creative at all: the algorithm simply choose what to write based on what was in the input corpus. There is no space for completely new elements: this is what Margaret Boden defined ``Combinatorial Creativity'', that is the lowest level of creativity. This justifies the fact that researchers did not stop looking for new methods for the generation of music. One most notable advancement is the use of machine learning algorithms, and in particular deep learning techniques, to generate less obvious and more novel reinterpretations of a corpus~\cite{briot_deep_2017}. Arguably, these techniques could in principle reach higher levels of creativity, as the results obtained by a machine learning software can surprise even his programmer. Against this there is the ``Chinese Room'' argument by Searle, who advocates that any algorithm, despite its complexity, cannot be really intelligent or creative because the machine cannot really understand what is being processed~\cite{searle_minds_1980}. Even considering this, that is beyond the point that is being made here: new algorithms are being proposed for music generation even if we already have some effective ones, because the researchers are trying to obtain results that can be considered \textit{more creative}.

This brings us back to the fundamental question, what does it mean to be creative? And, more pragmatically, how can one assess whether an algorithm is more or less creative? A scientist should always try to give measurable indications of how well the hypothesis of a work were met, and in this case it would mean to grade the creativity of a proposed algorithm. This is by no means easy, due to the difficulty of defining what is creative and to the nature of art's aesthetics, that is often considered subjective. Despite the difficulty of the task, a lot of publications propose methods to assess creativity~\cite{pease_evaluation_2018}.

\subsection{A Fragmented Scene}
From what I described, one could imagine that researchers in this field would try to stick to a definition of creativity, take the most promising works in literature, try to increase the level of creativity with their research, and evaluate their results accordingly. The reality is very different. Many proposals don't use previous works as a starting point, don't clearly state what is their goal in terms of creativity, and only a small fractions of the papers contain any reference to any kind of evaluation~\cite{jordanous_standardised_2012}. This is partly due to the fact that different professionals are involved, each having different aims: artists for instance might not be interested in giving creative behaviors to software, but would rather ``input'' their own creativity to obtain interesting results with the help of software tools. Even among researchers, and among those that do evaluate somehow their works, sometimes the goal is to obtain result that are ``valuable'' rather than creative, often meaning that the result is pleasing to listen to. 
In principle this is not wrong, but `floods' the scene of Musical Metacreation with works that do not really help the advancement of Computational Creativity, and in general that have little interest from a scientific point of view. It is not surprising that the scientific community has started to scorn ``mere generation'' systems~\cite{ventura_mere_2016}. For example, Huawei recently publicized how their AI chips were capable of finishing Schubert's eight Symphony\footnote{\url{http://consumer.huawei.com/uk/campaign/unfinishedsymphony.html}}. An impressive deed, but their method was to use a LSTM network to generate melodies, that a composer then selected, harmonized and orchestrated, leaving very little creativity room to the artificial intelligence. 
On the other hand, it is clear that creativity without artistic value is of little use, and that the desire for the creation of something valuable is probably what motivated this research field in the first place. 

\section{Conclusions and Directions}
\label{sec:conclusions}

In this paper I briefly described some of the issues of Musical Metacreation seen as a scientific research field with the intent of finding out if and how computers can be musically creative. Others have already pointed out these issues in reviews and position papers~\cite{colton_computational_2012}. Many suggested more rigor in evaluating the results (and in rejecting papers that do not sufficiently evaluate their results)~\cite{jordanous_standardised_2012}, and hybridization of results was also suggested~\cite{fernandez_ai_2013}. 

Concerning hybridization, that is still uncommon in research, I would suggest to use the strongly hierarchical nature of music~\cite{simonetta_symbolic_2018,carnovalini_multilayered_2019} to hybridize methods in a top-down manner: the general structure or form of the musical piece could be generated with simpler algorithms like Markov Chains, and each segment could be then filled with different methods generating an harmonic progression and then the melodies. In such a framework, one could fine-tune the creative intervention: even with classical music masterworks it is often true that what makes the composition really stand out as creative is limited to a few choices, while the rest of the composition can follow common rules of the period. One could try to search for an optimal trade-off between deep learning and structured artificial intelligence, a division reminiscent of Pascal's \textit{esprit de g\'eom\'etrie} and \textit{esprit de finesse}~\cite{pascal_pensees_1852}.

I mean to design a modular framework for music generation that should make the hybridization easier, and could possibly allow more effective comparisons between existing algorithms. Hopefully, this should make evaluations somewhat easier, and could allow researchers to focus on specific aspects of creativity, and on what is needed to allow a creative deed. I will try as well to investigate how to express emotions through music, and how a computer should choose what to express. This has a double aim: the first is related to music therapy applications of automatic music generation, and the second is to investigate whether humans perceive as more creative something they can more easily relate to in an emotional way.